\documentclass[floatfix,prl,twocolumn]{revtex4}
\usepackage{keyval} \usepackage{dcolumn}

\usepackage[]{amsmath}
\usepackage[psamsfonts]{amsfonts}
\usepackage[psamsfonts]{amssymb}

\usepackage{graphicx}
\usepackage{bm}

\usepackage[usenames,dvipsnames]{color}

\newcommand{\threej}[6]{\ensuremath{\left(\begin{array}{ccc}#1 &#2 &#3 \\ #4 &#5 &#6 \end{array}\right)}}

\begin{document}
\title{The low spin moment in LaOFeAs is due to a hidden multipole order caused by spin orbital ordering.}
\author{Francesco Cricchio, Oscar Gr{\aa}n\"as \& Lars Nordstr\"om}
 \address
 {Department of Physics and Materials Science, Uppsala University, Box 530, 751 21,
Uppsala, Sweden}

\begin{abstract}
{An antiferro-magnetic (AF)  low moment solution, 0.4 $\mu_\mathrm{B}$/Fe, is found  in the case of LaOFeAs for an intermediately strong Coulomb interaction $U$ of 2.5--3.0 eV.  This solution is stabilized over a large moment solution due to the gain in exchange energy in the formation of large multipoles of the spin magnetization density.  The multipoles are of rank four and can be understood as a kind of spin-orbital ordering. Parallels can be drawn to the stabilization of the AF order in e.g.~CaCuO$_2$.
}
\end{abstract}

\maketitle
With the discovery of the iron pnictide layered superconductors in 2008 \cite{Kamihara:2008p334}, 
a hope was quickly raised that these materials would finally lead to an understanding of the elusive mechanism of the superconductivity of
the high-$T_\mathrm{C}$ cuprates.
Indeed there are many common features; the fact that the parent compound is anti-ferromagnetic (AF), the
central role played by a transition metal layer, the fact that the AF order quickly disappears with doping and then is overtaken by a strong superconducting state.
However, fairly soon some differences were also discovered. 
While the main electrons in the cuprates are correlated and close to an insulating state, in the iron pnictides they seems to be at most moderately correlated and metallic \cite{Norman:2008p17087,Tesanovic:2009p16813}.  
This difference between the two types of materials is also manifested  by the fact that density functional theory (DFT) based calculations of the undoped iron pnictides obtain the correct metallic AF order while in the undoped cuprates they falsely lead to a non-magnetic metallic state.
This latter failure is due to the neglect of the strong correlation among the Cu $3d$ states which is believed to play a
crucial role in forming the superconducting state in the doped materials. 
For instance, if a correlation term is added to the DFT Hamiltonian, local density approximation plus added Coulomb $U$ interaction formalism (LDA+$U$), an AF insulating phase is obtained  \cite{Anisimov:1991p16790}. 
However, with the ever increasing number of DFT studies, it has been clarified that DFT has problems also for the iron pnictide parent compounds, although of different nature \cite{Mazin:2008p17085}.
The calculations systematically overestimate the ordered AF spin moment, which is 0.35 $\mu_\mathrm{B}$ in LaOFeAs \cite{Cruz:2008p16876}.
In fact, state-of-the-art DFT calculations in the generalized gradient approximation (GGA) give spin moments of the order 2.0--2.5 $\mu_\mathrm{B}$ \cite{Mazin:2008p17085,Opahle:2009p16846}, i.e.~an overestimation by at least a factor five.

In this Letter we perform LDA+$U$ calculation for the AF parent compound LaOFeAs.
The obtained results show, that for realistic
$U$ parameters, a low spin moment solution is stabilized due to polarization of higher multipole moments of the spin density. These terms can be analyzed as a spin orbital ordering among mainly the $xz$ and $yz$ $d$-orbitals at the Fe sites. 
Finally we make a comparison with the LDA+$U$ solution for an undoped cuprate, CaCuO$_{2}$,
which reveals a striking similarity in the role played by magnetic multipoles.

The electronic structure is calculated within the full-potential augmented plane wave plus local orbital (APW+lo) method as implemented in the {\sc elk} code \cite{elk}.
The LDA+$U$ approach is applied following the same methodology as described in Ref.~\onlinecite{Bultmark:2009p16779} with Yukawa screening \cite{MRNorman:1995p140}   and 
around mean-field (AMF) double counting
while 
the GGA \cite{Perdew:1996p17060} is used for the DFT part.
The  AFM Brillouin zone is sampled with $10\times10\times6$ $\vec{k}$ points  uniformly spaced. The calculations are done for the crystal parameters of the experimental high  temperature tetragonal structure \cite{Cruz:2008p16876}.
The dimension-less parameter governing  the number of augmented plane waves 
$R |\vec{G}+\vec{k}|_{\mathrm{max}}$ is chosen to be 8.0, where
$R$ is the Fe muffin tin radius and $\vec{G}$ are the reciprocal lattice vectors. 

There have been several attempt to estimate the magnitude of the Coulomb interaction $U$ in this compound. The results stretch all the way from fairly large values of 4 eV leading to strong correlation \cite{Haule:2008p17053}, through moderate values of 3-4 eV \cite{Anisimov:2009p17055} and 2.7 eV \cite{Aichhorn:2009p17054}, down to less than 2 eV \cite{Yang:2009p17073}. As has been discussed \cite{Anisimov:2009p17055,Aichhorn:2009p17054}, part of the disagreement stems from the different choices of band manifolds that are allowed to interact with this Coulomb interaction. If a down-folding down to a subset of Fe $d$-states is performed, the effective 
Coulomb interaction has to be decreased too, otherwise the effect of correlation is overestimated.
In the present study we will vary $U$ between 0 and 4 eV, where the 0 eV case corresponds to a pure GGA calculation, since all Slater parameters are screened with the same Yukawa screening parameter \cite{Bultmark:2009p16779}.
In this approach the Hund's rule exchange parameter $J$ varies automatically between 0 and 1 eV, with e.g.~$J$=0.82 eV for $U$=2.5 eV, a set of values which is very close to the values obtained by a constrained DFT approach \cite{Aichhorn:2009p17054} with $U$=2.7 and $J$=0.79 eV.

\begin{figure}[htb]
  \centering
 \includegraphics[width=0.7\columnwidth]{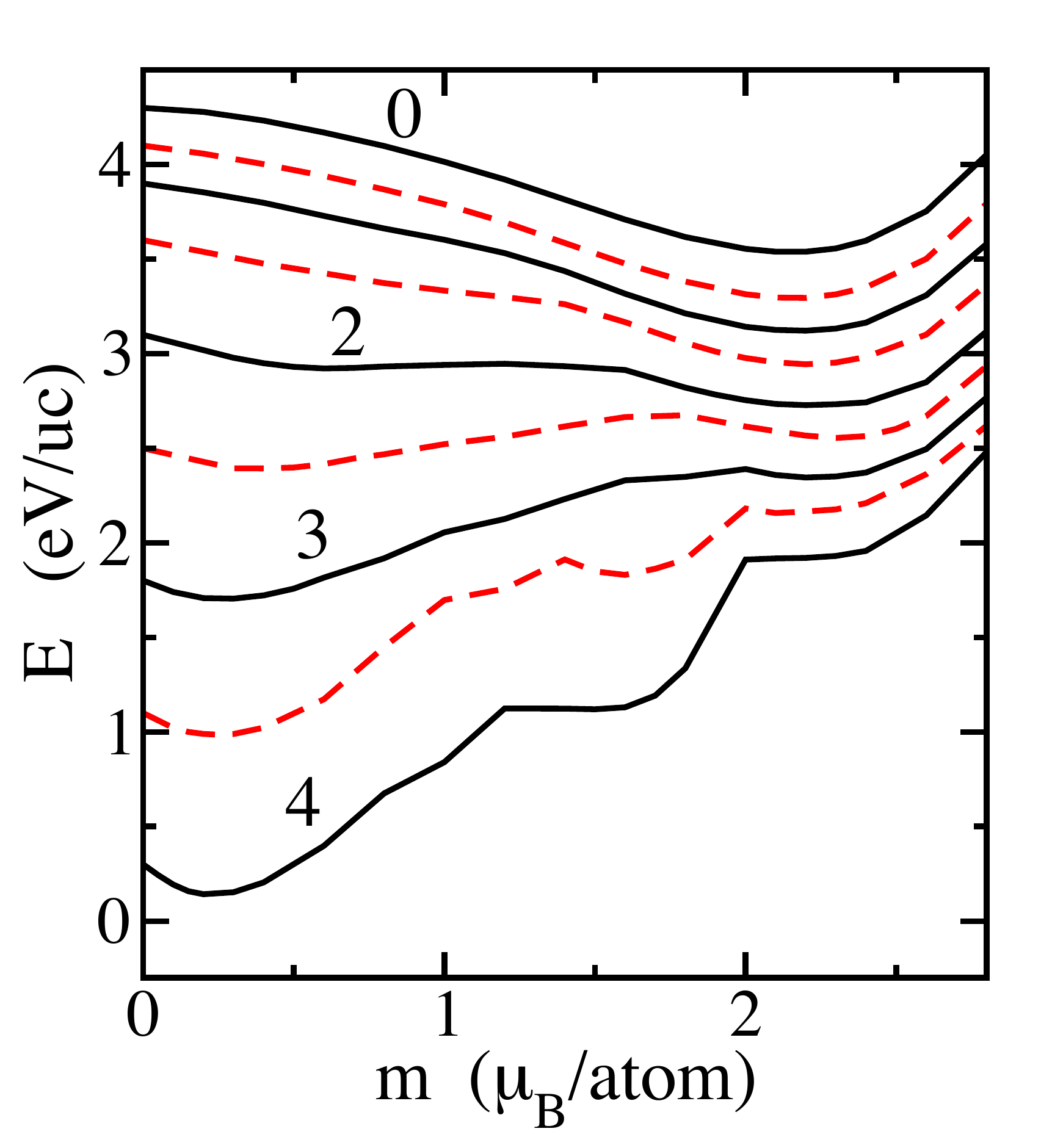}
  \caption{Total energy per magnetic unit cell (4 formula units) as a function of the staggered spin moment per Fe atom  calculated with varying $0\leq U \leq 4$ eV in steps of 0.5 eV (some values are indicated), with solid curves for integer values. Notice that the energy shifts between the curves are arbitrary and chosen such as to simplify the comparison. \label{EMU}}
\end{figure}

The total energy as a function of the spin moment, as obtained by constraining the staggered spin moments \cite{Dederichs:1984p13666} of the stripe ordered AF state, and as a function of $U$, is displayed in Fig.~\ref{EMU}. 
In agreement with earlier studies \cite{Mazin:2008p17085} the GGA curve ($U=0$) has a clear deep minimum at $m=2.2\,\mu_{\mathrm{B}}$. This minimum moves slightly to larger moments by increasing $U$. 
Simultaneously, a second solution starts to develop at a smaller moment. At $U\approx2$ eV this has evolved to a local minimum, which becomes the global minimum for $U\gtrsim2.5$ eV, a value close to the estimated one \cite{Aichhorn:2009p17054}.
At the largest values of the Coulomb parameter also an intermediate minimum is formed.
The stabilization energy of the low moment state is large. Already the high moment solution of GGA had a significant stabilization energy of 0.17 eV per formula unit (fu), for $U$=2.5 eV the low moment solution is lower than the high moment solution by 0.04 eV.
It is here worth noticing that except for the pure GGA calculation the $m=0$ solution is not stationary. 
This is an indication that $m=0$ generally is not a time reversal (TR) symmetric state. 
In fact it can be much lower in energy than the TR-symmetric solution, about 0.2 eV/fu for $U$=2.5 eV. 

\begin{figure}[htb]
  \centering
   \includegraphics[width=0.8\columnwidth]{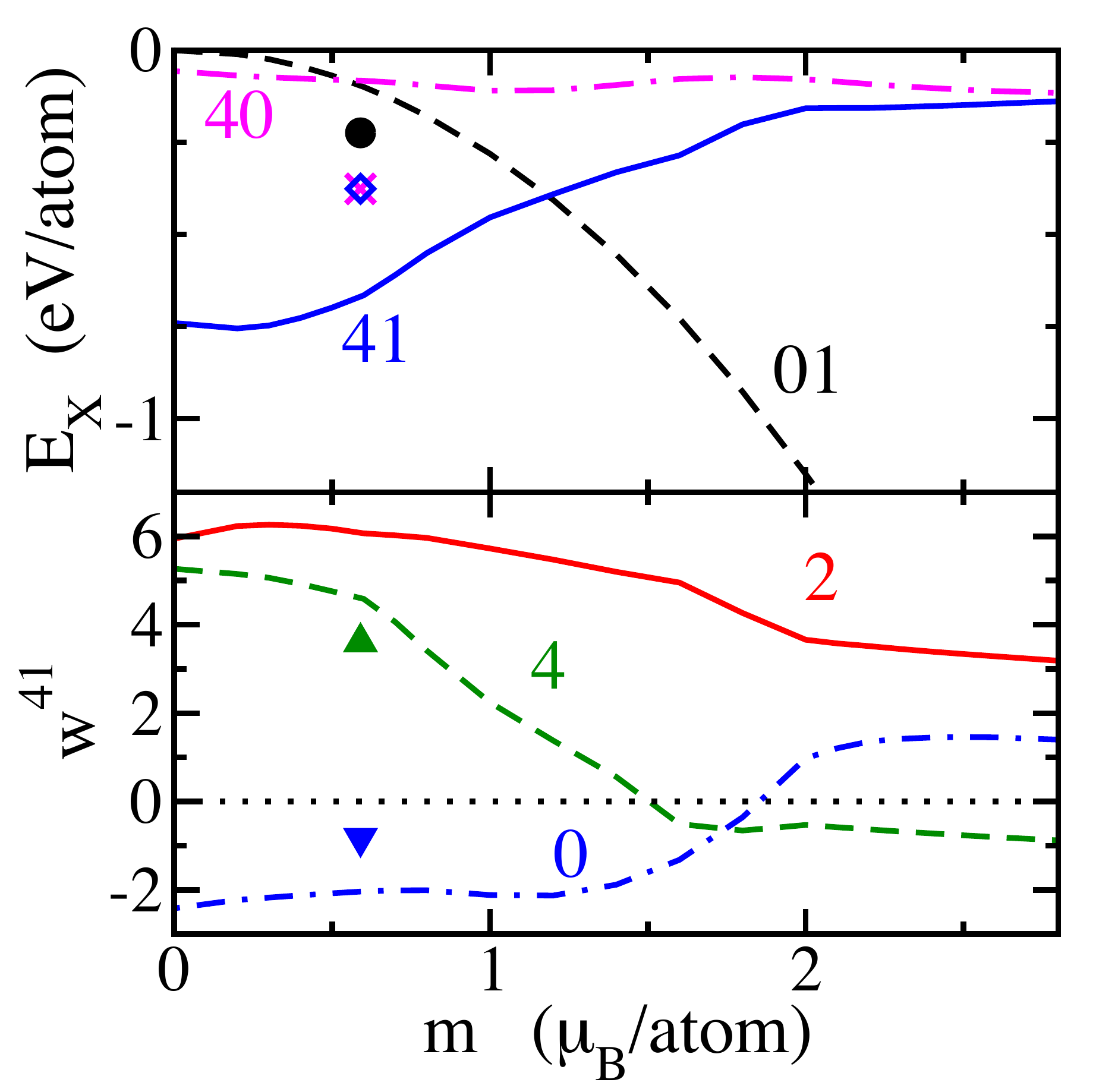}
  \caption{Exchange energy per Fe atom decomposed into multipole $\mathbf{w}^{kp}$ contributions (top, where the number indicates $kp$) and multipole tensor components $w^{41}_{q0}$ (bottom, with numbers indicating $q$) as a function of the spin moment per atom for a fixed $U$=2.5 eV. 
  In addition the same quantities are shown with symbols obtained by a corresponding calculation for CaCuO$_{2}$ with $U$=7.0 eV and calculated $m$=0.59 $\mu_\mathrm{B}$.
  \label{Ex-W}}
\end{figure}

In order to analyze this low moment solution, that is stabilized at physical values of the Coulomb parameter $U$, we will adopt the 
multipole tensor formalism which has been described in some detail earlier \cite{Bultmark:2009p16779}.
The multipole tensors can be obtained from the density matrix through \cite{Laan:1995p51,Bultmark:2009p16779}
\begin{align}
w^{kp}_{qt}\equiv{w}_\alpha
= \mathrm{Tr} \,{{\Gamma}}_\alpha\, {\rho}
\,,
\end{align}
where the matrix elements of the corresponding expansion matrices $\Gamma_\alpha$ are given by \cite{Cricchio:2008p1260,Bultmark:2009p16779}
\begin{align}
\Gamma^{kp}_{qt;ab}&\equiv\Gamma_{\alpha;ab}=N_{kp}^{-1} \,(-)^{m_{a}-\ell+s_{a}-s} \nonumber \\
&\times \mathcal{T}\threej{\ell}{k}{\ell}{-{m}_{a}}{q}{m_{b}}
\mathcal{T}
\threej{s}{p}{s}{-{s}_{a}}{t}{s_{b}}\,,
\end{align}
where $\alpha=\left\{kp;qt\right\}$ is a composite index for the double tensor indices $k$ and $p$ and the corresponding components $q$ and $t$, the $\left(\dots\right)$-symbol is the Wigner-3j symbol, and $N_{kp}=n_{\ell k}n_{sp}$, where $n$ is the usual normalization factor \cite{Laan:1995p51,Bultmark:2009p16779}.
The operator $\mathcal{T}$ transforms the spherical tensor, which was used in earlier studies \cite{Cricchio:2008p1260}, to a tesseral form \cite{tesseral}. 
In this study 
we prefer to work with tesseral tensor moments $\mathbf{w}^{kp}$  since 
they ensure that the matrices $\Gamma_\alpha$ are all hermitian and hence simpler to interpret.
The interpretation of these multipole tensors are that for even $k$ they correspond to the multipoles of the charge ($p$=0) or spin magnetization ($p$=1), while for odd $k$ they are multipoles of the corresponding currents.

The screened exchange energy takes a very simple and appealing form when expressed in terms of these tensor moments 
\cite{Cricchio:2008p1260,Bultmark:2009p16779},
\begin{align}
E_{\mathrm{X}}=
-\frac{1}{4}\sum_\alpha A_{k} w_\alpha^{2}\,.\label{Ex-w}
\end{align}
Firstly, one note the resemblance with the Stoner  exchange, $-Im^{2}/4$, and indeed one can identify $I=A_{0}$ since $m=w^{01}_{00}$. 
Secondly, all other multipole moments contribute to the exchange energy in the exact same way; with an energy parameter times the multipole moment squared.

\begin{figure}[htb]
  \centering
           \includegraphics[width=0.8\columnwidth]{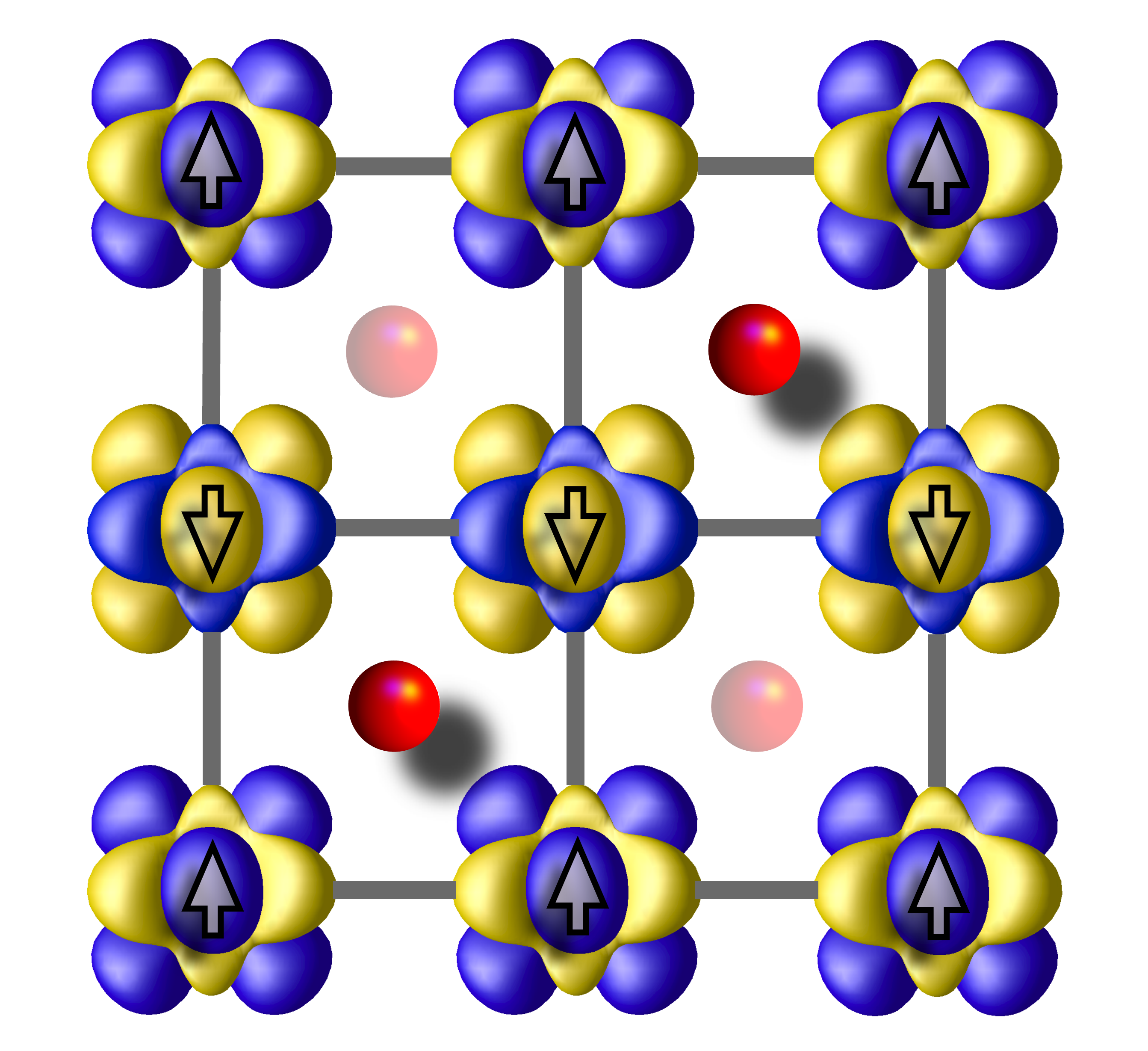}
  \caption{
 Isosurface plots of  the magnetization density around the Fe sites for the  striped AF order and $U$=3.0 eV are displayed with positive value indicated with dark/blue and negative with light/yellow. The arrows show the directions of the small integrated atomic dipole moments. The As atoms situated below and above the Fe-plane are displayed with spheres.\label{m-plot}}
\end{figure}

The contribution from the exchange energy to the total energy curve  for $U$=2.5 eV of Fig.~\ref{EMU}, is decomposed in the contribution from the multipoles $\mathbf{w}^{kp}$
according to Eq.~(\ref{Ex-w}) 
and displayed in Fig.~\ref{Ex-W}.
Besides the spin polarization energy, which is of course quadratic with the moment, there is a large exchange contribution from the magnetic multipole $\mathbf{w}^{41}$. 
Since it has the largest magnitude for small moments where it dominates, it is the one that stabilizes the small and intermediate moment solutions for large enough $U$ in Fig.~\ref{EMU}.
The most significant multipole tensor components $w_{\alpha}$ as a function of the constrained moment are also displayed in Fig.~\ref{EMU}.
There exist three independent  components of $\mathbf{w}^{41}$:  $q=0$, 2 and 4. They are the symmetry allowed hexadecapoles (rank four) of the spin magnetization density which are rotationally invariant, two-fold invariant, and four-fold invariant, respectively, around a tetragonal axis through a Fe site. While $q$=0 and 4 are both allowed also for a local tetragonal symmetry, the $q$=2 is permissible due to lower symmetry at the individual Fe sites caused by the striped AF order. These large multipoles result in a very anisotropic magnetization density as seen in Fig.~\ref{m-plot} for the case of $U$=3.0 eV,
where the magnetization density has both large positive and negative values but integrates to a small value of 0.3 $\mu_{\mathrm B}$.

This low moment solution is stabilized through an intricate competition between gain in screened exchange energy and loss in kinetic energy. The gain in exchange energy by the Coulomb interaction of Eq.~\ref{Ex-w}
is manifested in the orbital dependent exchange potential matrix which enters the LDA+$U$ hamiltonian, 
\begin{align}
V_{\mathrm{X}}=\frac{\partial E_{\mathrm{X}}}{\partial \rho^\mathrm{T}}=
\sum_\alpha\frac{\partial E_{\mathrm{X}}}{\partial w_\alpha}\frac{\partial w_\alpha}{\partial \rho^\mathrm{T}}
=-\frac{1}{2}\sum_\alpha A_{k} {w_\alpha}\,\Gamma_{\alpha}\,.\label{Vx-w}
\end{align}
Again, since $\Gamma^{01}_{00}=1\otimes\sigma_{z}$, it is possible to identify the Stoner exchange splitting $\Delta_{\mathrm S}=Im= A_{0}w^{01}_{00}$, that can be generalized to  the higher multipole splitting  $\Delta_{\alpha}=A_{k} w_{\alpha}$.
The corresponding $\Gamma_{\alpha}$ matrix describes which kind of states will split due the multipole $\alpha$ and generally involves the orbital degrees of freedom.
The $\Gamma$ matrix for the most significant component of the magnetic multipole, in the orbital basis of $xz$, $yz$, $xy$, $x^{2}-y^{2}$ and $z^{2}$, is given by
\begin{align}
\Gamma^{41}_{20}=\left(
\begin{array}{ccccc}
 2 \sqrt{5} & 0 & 0 & 0 & 0 \\
 0 & -2 \sqrt{5} & 0 & 0 & 0 \\
 0 & 0 & 0 & 0 & 0 \\
 0 & 0 & 0 & 0 & \sqrt{15} \\
 0 & 0 & 0 & \sqrt{15} & 0
\end{array}
\right)\otimes\sigma_{z}\label{mat-20}\,,
\end{align}
where $\sigma_{z}$ is the Pauli spin matrix. Hence, the existence of $w^{41}_{20}$ moments manifests an ordering of spin-orbitals, by e.g.~a spin dependent splitting of the $xz$ and $yz$ Fe-$d$ orbitals.
Similar spin-orbital orderings has been recently suggested to play a role in  LaOFeAs
\cite{Kruger:2009p17086}.

\begin{figure}[htb]
  \centering
   \includegraphics[width=1\columnwidth]{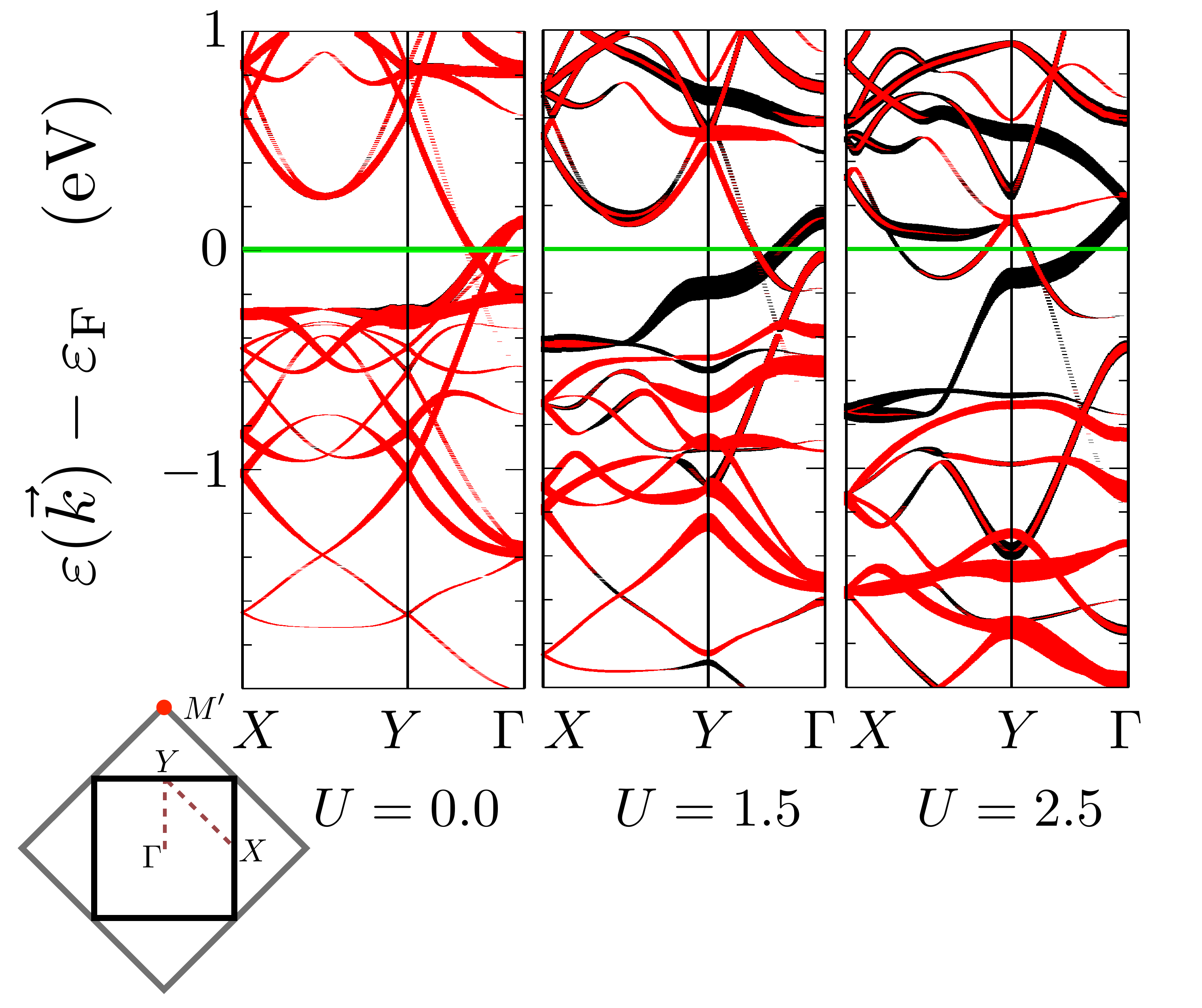}
  \caption{The band structure $\varepsilon(\vec{k})$ along the path $X$-$Y$-$\Gamma$ in the Brillouin zone of the striped AF order (shown in the inset together with the larger non-magnetic BZ with the corresponding wave vector of the magnetic order $M'$)  for varying Coulomb parameter $U$ and fixed local Fe moment of 0.4 $\mu_{\mathrm B}$. The zero of energy is at the Fermi energy $\varepsilon_\mathrm{F}$.
  The width of the bands are proportional to the expectation value of $\Gamma^{41}_{20}$ for the band states, 
  with the dark/black (light/red) color specifying a negative (positive) value.
   \label{band}  }
\end{figure}

In 
Fig.~\ref{band} the band structures are displayed  for three different values of $U$ with the local staggered magnetic moment constrained  to a value representative to low moment solution, 0.4 $\mu_\mathrm{B}$. The $\vec{k}$-path is in the basal plane of the Brillouin zone (BZ) starts at $X$, goes via $Y$, which is half the magnetic ordering wave vector $M'$, and ends at $\Gamma$ in the center of the BZ (see the inset of Fig.~\ref{band}). 
The two points $X$ and $Y$ become equivalent in the case of TR symmetry.
Firstly, we will take a look at the GGA ($U$=0) solutions. 
One can see the nesting features of the bands at the Fermi energy along the $Y\Gamma$-line. These bands originates from the hole Fermi surface originating from the $\Gamma$ point and the electron Fermi surfaces down-folded from the $M'$-point of the larger TR symmetric BZ \cite{Singh:2008p17118}. Surprisingly, these bands seem to be inert to the nesting effect, since the bands cross also in the AF structure with no hybridization gap opening up.
This is due to the different orbital character of these nested bands. 

Now, we will focus on the stabilization of the low moment solution due to the formation of multipoles. For  finite $U$ one can see that the bands become polarized through the action of the orbital dependent potential of Eq.~(\ref{Vx-w}), as illustrated by the ``fatness'' and color of the plotted bands in Fig.~\ref{band}. The width is proportional to the magnitude of the expectation value 
of the matrix $\Gamma_{20}^{41}$ with light and dark colors indicating positive and negative values. Here we can see that  there are large splittings due to the polarization of the $w_{20}^{41}$ tensor components, which results in strong rearrangements, especially along the $Y\Gamma$-line. This produces a large  asymmetry along the $XY$-line.
This splitting together with the one of the $w_{40}^{41}$ tensor components leads to an effective opening of a pseudo-gap at the the Fermi energy and a stabilization of this low moment solution. For larger $U>$4 eV, this effect is so strong that a true gap opens up and the solution becomes insulating. 
Again, as is clear from the intermediate $U$=1.5 eV plot, the Fermi surface nesting has no direct role in the stabilization of this low moment solution, which is in agreement with a recent angular resolved photo-emission spectroscopy experiment \cite{Liu:2009p17108}.

Finally, we want underline the fact that the multipole needed to stabilize the low moment solution, instead of the large moment solution as predicted by GGA, also play a crucial role in the formation of an insulating AF solution in the cuprates, as e.g.~CaCuO$_{2}$. The multipoles and their energies for CaCuO$_{2}$ calculated with $U$=7.0 eV are shown in Fig.~\ref{Ex-W}. In this case the existence of the multipole is easier to understand as it is essentially a pure $x^{2}-y^{2}$ orbital that polarizes, giving arise to a non-spherical charge and magnetization density. 
However, the magnitude of the multi-poles are of the same order as in LaOFeAs, and in fact
without these multipoles, the non-magnetic solution is more stable. The last is accordance with the fact that more exchange energy goes into the formation of the multipole than that of the spin moment.
Hence, in both types of compounds it is the neglect of these multipole exchange channels in LDA and GGA that lead to the wrong ground state, with either too large (LaOFeAs) or too small moments (CaCuO$_{2}$). This favorable comparison between the magnetism of the undoped LaOFeAs and an undoped cuprate will be explored in further details  in a future study. Then a crucial issue remains wide open; how do these spin and spin-orbital ordered AFM ground states of the parent compound, with their significant formation energies, vanish already with a small doping, which eventually leads to a high $T_\mathrm{C}$ superconductivity? One can speculate that  the multipole order in some form remains beyond the doping where the AFM order is destroyed, and then constitutes the hidden order of the so-called pseudogap region, which is well established in the cuprates \cite{Timusk:1999p17136} and has been observed recently for the pnictides \cite{Kato:2009p17117}.

The support from the Swedish Research Council (VR) is thankfully acknowledged. The computer calculations have been performed at the Swedish high performance center HPC2N 
under grant provided by the Swedish National Infrastructure for Computing (SNIC).


\begin{thebibliography}{24}
\bibitem[{tes()}]{tesseral}
\bibinfo{note}{A tesseral tensor $x^{r}$ is related to the spherical tensor
  $\tilde{x}^{r}$ through ${x}^{r}_{0}=\tilde{x}^{r}_{0}$ and
  ${x}^{r}_{q}=\left(\tilde{x}^{r}_{|q|}+(-)^q \tau
  \tilde{x}^{r}_{-|q|}\right)/\sqrt{2\tau}$ for $q\ne0$, where
  $\tau=\mathrm{sgn}\,q$.}
\end{thebibliography}
\end{document}